\documentstyle[aps]{revtex} \draft \begin{document} \title{Kramers-like
picture for crystal nucleation }
\author{D.  Reguera, J.M.  Rub\'{\i}, A.  P\'erez-Madrid} \address{Departament 
de F\'{\i}sica Fonamental\\
Facultat de F\'{\i}sica\\ Universitat de Barcelona\\ Diagonal 647, 08028
Barcelona, Spain\\ }

\maketitle

\begin{abstract}

We introduce a new scheme to analyze the kinetics of homogeneous nucleation in 
terms of a global order parameter. Our approach is based on the application of 
the internal degrees of freedom formalism$^{\ref{bib:degroot}}$ to derive a kinetic
 equation of the Kramers 
type formulated for a global reaction 
coordinate. We provide explicit  expressions for the quantities and coefficients involved in the 
process, suitable for simulation.
 In addition, our picture recovers in the  
quasi-stationary case the transition rate obtained from the method of 
reactive flux.
The equation we present may provide a link between theoretical 
approaches to 
homogeneous nucleation (generally formulated in terms of a kinetic equation of 
the Fokker-Planck type) and simulations (which mostly employ linear response 
theory). In this context, our scheme provides a theoretical framework to 
interpret and 
extend the results obtained in recent
 simulations.$^{\ref{bib:frenkel}-\ref{bib:frenkel3}}$

\end{abstract}

\section{Introduction}

Nucleation is the fundamental mechanism of many phase transformations, and has 
been the subject of extensive theoretical and experimental 
investigations.$^{\ref{bib:reviews}}$
In the last years, computer simulations have also become a useful tool to study 
the kinetics of crystal nucleation. However, in order to analyze the results of simulations the most common theoretical treatments for nucleation processes are not generally employed.

The kinetics of homogeneous nucleation is habitually described from a theoretical point 
of
view by means of kinetic equations of the Fokker-Planck type focused on the 
size
distribution of the clusters present in the system.  These equations are
formally analogous to the ones arising from Kramers reaction-rate 
theory,$^{\ref{bib:kramers},\ref{bib:hanggi}}$ since
in fact nucleation is an activated process.  The several approaches proposed to 
evaluate
nucleation rates differ basically in the expressions for the drift and
diffusion coefficients of the Fokker-Planck equation.  The values of these
coefficients  are sometimes postulated, assumed as
unspecified parameters, or left in terms of the addition and loss rates
of one monomer to a cluster.  To estimate these rates a proper microscopic
kinetic model is required.  These theories are customarily used to evaluate
steady-state nucleation rates, and most of them assume the constancy of the
 diffusion coefficient.

Contrarily, two different ways to proceed exist for simulations.  On the one
hand, there is a type of simulation,$^{\ref{bib:simula}}$ which can be called
 "straightforward",
whose method consist of simply supercooling the system and then waiting for the 
appearance of
nuclei.  Since the nucleation barrier decreases as $1/ (\Delta T)^{2}$, where 
$\Delta
T$ is the degree of supercooling, this type of simulation based on observing the
spontaneous formation of nuclei requires experimentally inaccessible
supercoolings.  For more moderate and realistic conditions the nucleation
barrier is so large that a spontaneous crossing is extremely unlikely during the
typical times of a simulation.

Recently, a different type of simulation, allowing the study of 
nucleation under more
realistic undercoolings, has been introduced by ten Wolde, Ruiz-Montero 
and Frenkel in refs. \ref{bib:frenkel}-\ref{bib:frenkel3}. Their
method is based on the separation of the simulation into two parts:  the 
evaluation
of the nucleation barrier following the Van Duijneveldt-Frenkel 
scheme$^{\ref{bib:barrera}}$ and the subsequent
calculation of the nucleation rate by means of the Bennett-Chandler
 picture,$^{\ref{bib:chandler}-\ref{bib:bennett}}$ based on
linear response theory.
The macroscopic state of the simulated system is described by means of a 
reaction
coordinate connecting the initial and final phases and sensitive to the
global degree of crystallinity in the system.  Concretely, Frenkel and
co-workers use a set of bond-order parameters introduced by Steinhardt 
$et\;al.$$^{\ref{bib:param}}$

One may wonder why they use linear response theory instead of appealing to the
usual nucleation theories or even to the prototypical theory of activated
processes:  the Kramers theory. In fact, in their 
simulations$^{\ref{bib:frenkel},\ref{bib:frenkel2}}$ they report on a diffusive behavior that seems to corroborate
the validity of Kramers picture. Additionally, they also employ Kramers
equation in order to obtain reasonable estimates of the time evolution of the
system.

In reference {\ref{bib:frenkel4} the authors themselves give the reasons why 
they do not use
Kramers theory.  First, they assert that it was
originally formulated for the single particle diffusion problem and therefore
its validity is
not guaranteed in the case when the reaction coordinate is a global one.  On the
other hand, they point out that the diffusion coefficient, assumed to be 
constant
 in Kramers picture,
may depend on the value of the reaction coordinate, and moreover its expression
may be non-trivial.  Finally, they affirm that a Fokker-Planck description of
nucleation processes does not take into account hydrodynamic effects that may be
very important in this type of barrier-crossing processes.

Our purpose in this paper is to introduce a theoretical description of 
homogeneous nucleation in
terms of a global order parameter like the one used by Frenkel and co-workers in
their simulations. The treatment we present is based on the 
kinetic equation 
introduced in a previous paper$^{\ref{bib:yo}}$ and derived from the application of
 the internal degrees of
freedom formalism.$^{\ref{bib:degroot},\ref{bib:prigo}}$  We want to show how the 
scheme we propose provides a complete
description of the process, it includes the transition rate expression of the 
method of reactive flux  
as a quasi-stationary limit, and it overcomes the reservations exposed by 
Frenkel $et\;al.$

The paper is structured as follows.  In section II, we 
outline and briefly review the method of reactive flux from which expressions
used by Frenkel's group in their simulations are derived.  In section III, we 
introduce and extend the internal degrees of freedom formalism to the 
situation where the
internal coordinate is a non-local order parameter.  Within this framework, we
 derive a Kramers-like equation, which in section IV is
particularized to the quasi-stationary case to prove that it recovers the 
results
of the method of reactive flux.  In section V, we describe the way to proceed
beyond the quasi-stationary limit.  Finally, in the last section we 
summarize
and comment on the main results of the theory we propose.

\section{ Method of reactive flux.}

Our aim in this section is to present a brief review of the derivation, 
the
limitations, and the scope of applicability of the method of reactive flux to
evaluate reaction rates. To this end,
let us consider a macroscopic system whose state is characterized by a
reaction coordinate $Q$.  This system can be in two different states, $A$ and
$B$, separated by an energy barrier whose maximum is located at position 
$Q_{0}$. 
 The
state of the system, i.e. the side of the barrier where it is located,
 can be characterized by the functions

\begin{equation}\label{eq:nat} n_{A}(t)\,=\,\theta [ Q_{0}\,-\,Q(t) ]
\end{equation}

\begin{equation}\label{eq:nbt} n_{B}(t)\,=\,\theta [ Q(t)\,-\,Q_{0} ]
\end{equation}

\noindent where $\theta$ is the step function, and $n_{A}(t)$, $n_{B}(t)$ 
obviously satisfy the condition

\begin{equation}\label{eq:identidad} n_{A}(t)\,+\, n_{B}(t)\,=\,1
 \end{equation}

 \noindent  The probability that our system
is in state $A$ at time $t$ will be given by the non-equilibrium average
of the characteristic function $n_{A}(t)$

\begin{equation}\label{eq:pat} P_{A}(t)\,=\,\langle n_{A}(t) \rangle
\end{equation}

The starting point of this theory is to postulate the validity of the following
phenomenological equation describing the dynamics of the populations in states
$A$ and $B$

\begin{equation}\label{eq:pheno} \frac{d
P_{A}(t)}{dt}\,=\,k_{BA}P_{B}(t)\,-\,k_{AB}P_{A}(t) \end{equation}

\noindent where $P_{B}(t)\,=\,1\,-\,P_{A}(t)$, and $k_{AB}$ and $k_{BA}$ are 
the forward and backward rates,
respectively.  These rates are assumed to be constant and satisfy the detailed
balance condition

\begin{equation}\label{eq:detail} \frac{k_{AB}}{k_{BA}}\,=\,\frac{ P_{B}^{eq}}
{P_{A}^{eq}} \end{equation}

In the spirit of Onsager's regression hypothesis, we assume that the
relaxation of $\Delta P_{A}(t)\,\equiv\,P_{A}(t)\,-\,P_{A}^{eq}$ from an initial
non-equilibrium deviation $\Delta P_{A}(0)$ follows the same exponential
decay as the equilibrium correlation function of the fluctuations

\begin{equation}\label{eq:decay} \frac{\Delta P_{A}(t)}{\Delta
P_{A}(0)}\,=\,\frac{ \langle \Delta n_{A}(0) \Delta n_{A}(t) \rangle_{eq}
}{\langle \Delta n_{A}(0) \Delta n_{A}(0) \rangle_{eq}}\,=\, e^{-\lambda t} \end{equation}

\noindent where $\Delta n_{A}(t)\,=\,n_{A}(t)\,-\,\langle n_{A} \rangle_{eq}$, 
the symbol $\langle \ldots 
\rangle
_{eq}$ represents an equilibrium average, and the relaxation  rate $\lambda$ is given by

\begin{equation}\label{eq:lambda} 
\lambda\,=\,k_{AB}\,+\,k_{BA} 
	\end{equation}

From these equations one can obtain the following formula for the transition
rate, employed by ten Wolde, Ruiz-Montero and Frenkel in their 
simulations$^{\ref{bib:frenkel}-\ref{bib:frenkel3}}$

\begin{equation}\label{eq:kab1} 
k_{AB}\,=\,\frac{ \langle  \dot{n}_{B}(0)
{n}_{B}(t) \rangle_{eq}}{\langle n_{A}\rangle_{eq}}
\,=\,\frac{ \langle  \dot{Q}(0)\delta[Q(0)-Q_{0}]\theta[Q(t)-Q_{0}]
\rangle_{eq}}{\langle n_{A}\rangle_{eq}}
\end{equation}

\noindent This expression is not valid for very short times (because the
relaxation of the system cannot be exponential at $t=0$).  Additionally, it
 is also
restricted to times $t$ satisfying the condition $t<< 1/ \lambda$, since during 
the derivation it is assumed that $e^{- \lambda t}\approx 1$.
Moreover, we notice that although $k_{AB}$ is a time independent rate, in 
expression (\ref{eq:kab1}) it is equated to a correlation function 
depending explicitly on time.  Therefore, the latter equation is only correct for 
times long enough that the correlations have reached a plateau value.

On the other hand, taking into account the identity

\begin{equation}\label{eq:nat2} n_{B}(t)\,-\, n_{B}(0)\,=\,\int_{0}^{t} dt'
\frac{ d n_{B}(t')}{dt'} \end{equation}

\noindent and since the equilibrium average $\langle \dot{n}_{B} {n}_{B}(0)\rangle_{eq}$
vanishes, one can rewrite eq.  (\ref{eq:kab1}) as follows

\begin{equation}\label{eq:kab2} k_{AB}\,=\,\int_{0}^{t}\frac{ \langle
\dot{n}_{B}(0) \dot{n}_{B}(t') \rangle_{eq}}{\langle n_{A}\rangle_{eq}}dt'
\end{equation}

As asserted before, this expression is only valid for times long enough that
the correlations have achieved a plateau.  For such values of the time, it is then
meaningful to assume that the velocity autocorrelation function has 
decayed to zero,
so that we can replace the upper limit of the previous integral by infinity.
Therefore, the final expression for the transition rate yields

\begin{equation}\label{eq:kabchandler} k_{AB}\,=\,\frac{1}{\langle
n_{A}\rangle_{eq}} \int_{0}^{\infty}\langle \dot{n}_{B}(0) \dot{n}_{B}(t)
\rangle_{eq}dt\,=\,\frac{1}{\langle
n_{A}\rangle_{eq}} \int_{0}^{\infty}\langle \dot{Q}(0) \delta[Q(0)-Q_{0}] 
\dot{n}_{B}(t)
\rangle_{eq}dt
\end{equation}

\noindent  This expression will be recovered by using the treatment we 
introduce in the next section.

\section{Non-equilibrium thermodynamics approach.  Internal degrees of freedom}

The key element of the approach we propose is the formulation of a
kinetic equation to
describe homogeneous nucleation in terms of the order parameter $Q$
characterizing the global degree of crystallinity of the system.
This equation has already been introduced in a previous paper,$^{\ref{bib:yo}}$
 and we will devote this section to an extended and complete derivation 
 of this
equation. The procedure is based on the application of
the internal degrees of freedom formalism, which constitutes a straightforward 
 method to obtain kinetic equations
for any adequate internal coordinate or degree of freedom of the system.
In particular, we use this formalism considering $Q$ as the
 system internal
coordinate. 

Until now, in all of previous papers where this formalism has been successfully 
applied,$^{\ref{bib:degroot},\ref{bib:prigo}-\ref{bib:gabriel}}$ the system was 
always described by 
means of a local internal coordinate identifying  each system
constituent independently. As a result, a Kramers-like equation was obtained.
However,  we are now interested in a kinetic description in terms of a single 
global order parameter $Q$, characterizing the whole system. Therefore, 
we need to extend the internal degrees of freedom formalism to allow a 
description 
in terms of a global internal coordinate. In fact, we face the same 
problem pointed out by Frenkel and co-workers$^{\ref{bib:frenkel4}}$ about  the 
validity and the necessity 
of an extension of Kramers picture when the reaction coordinate is a global 
one.

From the point of view adopted in previous 
papers,$^{\ref{bib:prigo}-\ref{bib:gabriel}}$ one worked with a system
of $N$ identical particles.  The internal coordinate characterized
the state of each particle, which evolved independently without
interacting with the remaining particles. The global state of the system was 
obtained at each
instant by means of a statistical recount of the individual states of the $N$
particles.

In our case, however, we are interested in a contracted description of the 
system based
on a single internal coordinate that characterizes it in a global way.
Proceeding by analogy, we consider $N$ system replicas, identical
in the sense that all of them are compatible with a given macroscopic state 
(identified for example by the total number of particles, the pressure, and
the temperature).  This set of replicas constitutes a non-equilibrium ensemble, 
defining the "system" we are going to deal with.  Since we are working
with an ensemble, we will be able to relate and obtain statistically the
properties of a single replica from the behavior of the whole ensemble. Denoting 
 the probability density by $\rho(Q,t)$, the
quantity $\rho(Q,t)dQ$ represents the fraction of replicas with a value of the
degree of crystallinity between $(Q,Q+dQ)$, at the instant $t$.  For $N$ large
enough, the latter expression could be interpreted as the probability that a
single replica of the ensemble has a value of the internal coordinate between
$(Q,Q+dQ)$.

Therefore, the value of a quantity $A$ evaluated for a single replica could be
assimilated with the expected value of this observable in the non-equilibrium
ensemble characterized by $\rho(Q,t)$.  That is

\begin{equation}\label{eq:Adete}  A(t)|_{replica}
\Longleftrightarrow \langle A (t)\rangle _{ensemble} \,=\, \int_{Q_{1}}^{Q_{2}}
A \rho(Q,t) dQ \end{equation}

Again, in order to derive our kinetic equation, our starting point will 
be the Gibbs equation for the entropy variation
associated with a diffusion process through the energy barrier that separates
the states $Q_{1}$ (the value of the order parameter in the metastable phase)
and $Q_{2}$ (corresponding to the new stable phase)

\begin{equation}\label{eq:dentropy} \delta
S\,=\,-\,\frac{1}{T}\int_{Q_{1}}^{Q_{2}}\mu(Q,t) \delta\rho(Q,t)dQ
\end{equation}

\noindent Here $\delta S$ is the variation of entropy per replica, $T$ is the
temperature (taken as constant) and $\mu(Q,t)$ is the conjugate chemical
potential.  The latter equation corresponds to the variation of entropy in a 
diffusion process, with the assumption that the internal energy and the 
volume of the system are constant.

It is worth pointing out that the Gibbs equation constitutes a 
reformulation of the Gibbs entropy postulate. Indeed, from the fact that we 
are working with system replicas that
obviously do not interact with each other, the expression for the chemical
potential is that for an ideal system

\begin{equation}\label{eq:muQ} \mu(Q,t)\,=\,k_{B}T\ln\rho(Q,t)\,+\,\Phi(Q)
\end{equation}

\noindent where $\Phi(Q)$ is the energy barrier to be surpassed by a replica in
order to achieve the new stable phase. Bearing in mind that at equilibrium the 
chemical potential becomes constant, the
following expression for the energy barrier could be obtained from the latter 
equation

\begin{equation}\label{eq:barriQ} \Phi(Q)\,=\,\mu_{eq}\,-\,k_{B}T\ln\rho_{eq}(Q)
 \end{equation}

\noindent Therefore, we can rewrite the chemical potential as follows

\begin{equation}\label{eq:muQalt} 
\mu(Q,t)\,=\,\mu_{eq}\,+\,k_{B}T\ln\frac{\rho(Q,t)}{\rho_{eq}(Q)}
\end{equation}

\noindent Substitution of this equation in expression (\ref{eq:dentropy}) 
yields

\begin{equation}\label{eq:dentropyGibbs} \delta
S\,=\,-\,k_{B}\int\delta\rho(Q,t) \ln {\frac{\rho}{\rho_{eq}}} dQ
\end{equation}

\noindent which is just the  expression for the variation of the entropy 
obtained from the Gibbs entropy postulate. Consequently, the equation we use as 
starting point in our derivation, namely eq. (\ref{eq:dentropy}), constitutes 
an alternative formulation of the 
Gibbs entropy postulate more proper for a non-equilibrium thermodynamics 
description. Therefore, it essentially represents the variation 
of $S$ with respect to its local 
equilibrium value in terms of the distribution over the possible internal 
states associated with different values of the order parameter $Q$.

The evolution of the number 
density $\rho(Q,t)$ is governed by the continuity
equation

\begin{equation}\label{eq:contiQ} \frac{\partial\rho(Q,t)}{\partial
t}\,=\,-\frac{\partial J(Q,t)}{\partial Q} \end{equation}

\noindent where $J(Q,t)$ is the diffusion current in the internal space.  From
eq.  (\ref{eq:dentropy}) we can obtain the entropy production associated with
 this diffusion process

\begin{equation}\label{eq:prodS} \sigma\,=\,-\frac{1}{T}\int
J(Q)\frac{\partial\mu(Q)}{\partial Q}dQ \end{equation}

\noindent where we have employed eq.  (\ref{eq:contiQ}) and a partial
integration has been performed, assuming vanishing flux at initial and final
states.

The entropy production (\ref{eq:prodS}) has the usual form of a sum of 
flux-force pairs. Following the non-equilibrium thermodynamics method, we may 
formulate the phenomenological equation

\begin{equation}\label{eq:leyfeno}
J(Q,t)\,=\,-\frac{L(Q)}{T}\frac{\partial\mu(Q,t)}{\partial Q} \end{equation}

\noindent where $L(Q)$ is the phenomenological coefficient, which may in 
general depend on the internal coordinate. To derive this expression, locality in 
the
internal space has been assumed, for which only fluxes and forces with 
the same value of the internal coordinate are coupled.

By substituting (\ref{eq:leyfeno}) in (\ref{eq:contiQ}) and using eq. 
(\ref{eq:muQ}) for the chemical potential, we obtain the following kinetic
equation in the internal space

\begin{equation}\label{eq:FPQ} \frac{\partial\rho}{\partial
t}\,=\,\frac{\partial}{\partial Q}\left( D(Q,t)\frac{\partial\rho}{\partial
Q}\,+\,b(Q,t)\frac{\partial\Phi}{\partial Q}\rho \right) \end{equation}

\noindent where

\begin{equation}\label{eq:b} b(Q,t)\,=\,\frac{L(Q)}{\rho(Q,t) T } \end{equation}

\noindent and

\begin{equation}\label{eq:D} D(Q,t)\,=\,k_{B}Tb(Q,t) \end{equation}

\noindent are the mobility and the diffusion coefficient in the internal space,
respectively.

Equation (\ref{eq:FPQ}) characterizes the evolution of the non-equilibrium
physical ensemble, determining in a complete way the nucleation process in
terms of the global order parameter $Q$.  In order to fully specify this
equation, we need an expression for the barrier $\Phi(Q)$ and for the
phenomenological coefficient $L(Q)$ relating the flux and its conjugate force 
in the internal space.

The value of the energy barrier is given by eq. (\ref{eq:barriQ}), 
and could be evaluated by means of simulations.  In fact,
this type of simulations for global order parameters involved in nucleation
processes has already been performed by van Duijneveldt and 
Frenkel.$^{\ref{bib:barrera}}$

On the other hand, to obtain an explicit expression for the phenomenological
coefficient $L(Q)$, we will adopt the fluctuating hydrodynamics 
framework.$^{\ref{bib:pago}}$  From
this point of view, one considers that the total current in the internal space
is the result of two contributions:  a systematic part $J^{s}$ and a random term
$J^{r}$ including the fluctuations inherent in the diffusion process along
the internal coordinate.  One has

\begin{equation}\label{eq:Joto} J(Q,t)\,=\,J^{s}(Q,t)\,+\,J^{r}(Q,t)
 \end{equation}

\noindent We will assume that $J^{r}$ has zero mean and satisfies the
fluctuation-dissipation theorem

\begin{equation}\label{eq:TFDQ} \langle J^{r}(Q,t)J^{r}(Q ',t ')
 \rangle\,=\,2k_{B}L(Q)\delta(Q - Q ' )\delta(t - t ') \end{equation}

\noindent Finally, from the latter equation one can derive the following
Green-Kubo expression for the phenomenological coefficient

\begin{equation}\label{eq:gkq} L(Q)\,=\,\frac{1}{k_{B}} \int_{Q_{1}}^{Q_{2}}d Q
 ' \int_{0}^{\infty}dt \langle J^{r}(Q',0)J^{r}(Q,t) \rangle \end{equation}

The objective of the two following sections will be to transform this
expression into a formula suitable for computer simulations.  First, we 
consider the
quasi-stationary case to show how to recover explicitly the method of reactive
flux expression commonly used to evaluate nucleation rates.  Subsequently, we
will analyze the general case.

\section{The transition rate for the quasi-stationary case.}

When the supercooling or the supersaturation is low enough, the energy barrier 
to
surpass in the nucleation process is large compared with thermal energy.  In
this high barrier situation, the system achieves a quasi-stationary state
characterized by a uniform current

\begin{equation}\label{eq:Jcuasi} J(Q,t)\,=\,J(t) \left\{ \theta(Q-Q_{1})\,-\,
\theta(Q-Q_{2}) \right\} \end{equation}

\noindent and a chemical potential which equilibrates independently at each 
side of the barrier

\begin{equation}\label{eq:mucuasi} \mu(Q,t)\,=\,\mu(Q_{1},t)\theta(Q_{0}-Q)\,+\,
\mu(Q_{2},t)\theta(Q-Q_{0}) \,\, ,\end{equation} 

\noindent where $Q_{0}$ represents the position of the top of the barrier and
$\theta$ is the step function.
By substituting this equation in (\ref{eq:muQ}) one obtains the following
expression for the probability density

\begin{equation}\label{eq:rhocuasi}
\rho(Q,t)\,=\,\rho(Q_{1},t)e^{-\frac{\Phi(Q)-\Phi(Q_{1})}{k_{B}T}}
\theta(Q_{0}-Q)\,+\,\rho(Q_{2},t)e^{-\frac{\Phi(Q)-\Phi(Q_{2})}{k_{B}T}}
\theta(Q-Q_{0}) \end{equation}

On the other hand, the diffusive current can be written in a more convenient way
as

\begin{equation}\label{eq:Jexp}
J(Q,t)\,=\,-D(Q,t)e^{-\Phi/k_{B}T}\frac{\partial}{\partial Q }e^{\mu/k_{B}T}
\end{equation}

\noindent By equating (\ref{eq:Jexp}) and (\ref{eq:Jcuasi}), and by integrating 
over $Q$
with the help of eq. (\ref{eq:mucuasi}), one arrives at the following
expression for the nucleation rate

\begin{equation}\label{eq:Jodete} J(t)\,=\,\frac{D(Q_{0})}{Q_{2}\,-\,Q_{1}}
e^{-\Phi_{0}/k_{B}T} \left ( e^{\mu_{1}/k_{B}T}\,-\,e^{\mu_{2}/k_{B}T} \right )
\end{equation}

\noindent where the subindexes $0$, $1$ and $2$ indicate that the corresponding
function is evaluated at $Q_{0}$, $Q_{1}$ and $Q_{2}$, respectively.

Having obtained the nucleation rate expression for the
quasi-stationary case,  our goal now
will be to show that it is equivalent to the method of reactive flux equation
derived from linear response theory.
An important difference between our formalism and the method followed in the 
simulations is that we are working with a non-equilibrium ensemble of replicas,
while in simulations one makes reference to a single system or a single replica.
The connecting point that links our scheme with the results referred to a single
system is the probabilistic interpretation of the ensemble number density
$\rho(Q,t)$.  

Bearing  this probabilistic interpretation in mind, we will see how to recover
the transition rate expression (\ref{eq:kabchandler}).  The first step will 
be to transform the continuity
equation into the phenomenological starting equation of the method of reactive 
flux. 
By
integrating eq. (\ref{eq:contiQ}) from $Q_{1}$ to $Q_{0}$, and using the
quasi-stationary condition (\ref{eq:Jcuasi}) and the result (\ref{eq:Jodete}) 
one obtains

  \begin{equation}\label{eq:depate} \frac{d}{dt}\int_{Q_{1}}^{Q_{0}}\rho(Q,t)dQ
\,=\,-J(t)\,=\,\frac{D(Q_{0})}{Q_{2}\,-\,Q_{1}} e^{-\Phi_{0}/k_{B}T} \left (
e^{\mu_{2}/k_{B}T}\,-\,e^{\mu_{1}/k_{B}T} \right ) \end{equation}

\noindent where $\int_{Q_{1}}^{Q_{0}}\rho(Q,t)dQ\,=\,P_{A}(t)$ is just the
probability that the system would be on the left hand side of the barrier, at time
$t$.

From (\ref{eq:b}) and (\ref{eq:D}) the diffusion coefficient expression at the
top of the barrier is given by

\begin{equation}\label{eq:DQ0}
 D(Q_{0},t)\,=\,\frac{k_{B}L(Q_{0})}{\rho(Q_{0},t)} \end{equation}

\noindent By using (\ref{eq:muQ}) and (\ref{eq:rhocuasi}), the density 
$\rho(Q_{0},t)$
could alternatively be expressed as

\begin{equation}\label{eq:rhoQ0}
 \rho(Q_{0},t)\,=\,e^{\mu_{1}/k_{B}T}e^{-\Phi_{0}/k_{B}T} \end{equation}

\noindent On the other hand, from eq.  (\ref{eq:rhocuasi}) and (\ref{eq:muQ}) 
it follows

\begin{equation}\label{eq:emu2}
 e^{\mu_{2}/k_{B}T}\,=\,\rho_{2}e^{\Phi_{2}/k_{B}T}\,=\, \frac
 {\int_{Q_{0}}^{Q_{2}}\rho(Q,t)dQ}{\int_{Q_{0}}^{Q_{2}}e^{-\Phi/k_{B}T}dQ} \,=\,
 \frac {P_{B}}{\int_{Q_{0}}^{Q_{2}}e^{-\Phi/k_{B}T}dQ} \end{equation}

\noindent Here $P_{B}$ represents the probability of being on the right hand 
side of
the barrier.  By introducing these two latter expressions into eq. 
(\ref{eq:depate}), one arrives at

\begin{equation}\label{eq:feno} \frac{d
P_{A}(t)}{dt}\,=\,k_{BA}(t)P_{B}(t)\,-\,k_{AB}(t)P_{A}(t) \end{equation}

\noindent where

\begin{equation}\label{eq:kab}
k_{AB}(t)\,=\,\frac{L(Q_{0})k_{B}}{(Q_{2}-Q_{1})P_{A}(t)} \end{equation}

\noindent and

\begin{equation}\label{eq:kba}
k_{BA}(t)\,=\,\frac{L(Q_{0})k_{B}}{(Q_{2}-Q_{1})P_{A}(t)}\frac
{\int_{Q_{1}}^{Q_{0}}e^{-\Phi/k_{B}T}dQ}{\int_{Q_{0}}^{Q_{2}}e^{-\Phi/k_{B}T}dQ}
\end{equation}

\noindent are the forward and backward rates, respectively.

Therefore, we have just obtained the equation governing the evolution of the
probability that the system would be on the metastable side of the barrier.  At
this point, we have to highlight two important differences with respect to the
method of reactive flux. First, we have obtained this equation as a particular 
case
(corresponding to a quasi-stationary situation) of a more general theory
of wider applicability. Moreover, the rate coefficients in our
expression are explicitly time-dependent, which does not happen in the 
formula
of the method of
reactive flux.  Finally, it is important to remark that our rate coefficients 
fulfill the
detailed balance condition

\begin{equation}\label{eq:balancedet} \frac{k_{AB}(t)}{k_{BA}(t)}\,=\,\frac{
P_{B}^{eq}} {P_{A}^{eq}} \end{equation}

\noindent as follows from eqs. (\ref{eq:barriQ}), (\ref{eq:kab}) and 
(\ref{eq:kba}). This condition guarantees that the flux vanishes at equilibrium.

The last task to carry out is to compare the rate $k_{AB}(t)$ with the reactive 
flux
formula (\ref{eq:kabchandler}).  Within that framework, the expression
(\ref{eq:kab1}) is identified with the rate $k_{AB}$ at times large enough so
that the correlations have achieved a stable value characterized by a plateau.
In our expression, for long times the probability $P_{A}$ will approach and can 
be replaced by  the
equilibrium value $P_{A}^{eq}\,=\,\langle
n_{A}\rangle_{eq}$, and therefore $k_{AB}$ will become time
independent.

On the other hand, the random component $J^{r}$ of the current is related to the fluctuations of
$P_{A}$ with respect to the equilibrium value.  Indeed, by integrating the
 continuity equation (\ref{eq:contiQ}) from $Q_{1}$ to
$Q$ one obtains

\begin{equation}\label{eq:contiN} \frac{\partial N(Q,t)}{\partial
t}\,=\,-J(Q,t)\,=\,-J^{s}(Q,t)\,-\,J^{r}(Q,t) \end{equation}

\noindent where

\begin{equation}\label{eq:N} N(Q,t)\,=\,\int_{Q_{1}}^{Q}\rho(Q',t)dQ'
 \end{equation}

\noindent Notice that eq. (\ref{eq:contiN}) is valid for arbitrary times. 
In a very short time scale -microscopic time scale$^{\ref{bib:landau}}$- we can write

\begin{equation}\label{eq:contideltaN} \frac{\partial \delta N(Q,t)}{\partial
t}\,=\,-J^{r}(Q,t) \end{equation}

\noindent where the time derivative $\frac{\partial \delta N(Q,t)}{\partial
t}$ is the instantaneous rate of change.$^{\ref{bib:zwanzing}}$ Therefore

\begin{equation}\label{eq:fluctudeltaN} \langle J^{r}(Q',0)J^{r}(Q,t) \rangle
_{eq} \,=\, \langle \frac{\partial \delta N(Q',0)}{\partial t} \frac{\partial
\delta N(Q,t)}{\partial t} \rangle _{eq} \end{equation}

\noindent with $\delta N\,=\,N\,-\,N_{eq}$.  In the quasi-stationary state,
 described by eq. (\ref{eq:Jcuasi}), the quantity 
 $\frac{\partial \delta N(Q',t)}{\partial t}$ is independent of the position 
 $Q'$ of the absorbing barrier. Hence, we can particularize for convenience the
  latter expression for $Q'=Q_{0}$, and taking into account that

\begin{equation}\label{eq:denA} \frac{d \delta N(Q_{0},t)}{dt}\,=\,\frac{d
}{dt}\langle \Delta n_{A} (t) \rangle \,=\,- \frac{d
}{dt}\langle \Delta n_{B} (t) \rangle 
\end{equation}

\noindent we can then rewrite the phenomenological coefficient
(\ref{eq:gkq}) as

\begin{equation}\label{eq:LQ0cuasi} k_{B}L(Q_{0})\,=\,(Q_{2}-Q_{1})
 \int_{0}^{\infty}dt \langle { \langle \dot{n}_{B}(0) \rangle \langle
 \dot{n}_{B}(t) \rangle} \rangle_{eq} \end{equation}

\noindent Finally, substitution of this expression in eq.  (\ref{eq:kab}) yields 

\begin{equation}\label{eq:kabfin} 
k_{AB}\,=\,\frac{1}{\langle
n_{A}\rangle_{eq}}\int_{0}^{\infty}dt \langle { \langle \dot{n}_{B}(0) \rangle
\langle \dot{n}_{B}(t) \rangle} \rangle_{eq} 
\end{equation}

\noindent By remembering the link between ensemble averages and the values
associated with a single system, the equivalence between this equation and
expression (\ref{eq:kabchandler}) is then obviously proven.

\section{General case}

The method of reactive flux used by Frenkel and co-workers in their simulations
considers that the barrier separates the system into two states, $A$ and $B$, 
and it is restricted to the evaluation of the transition rate between these
initial and final states. In particular, it provides no information about the
evolution of the intermediate configurations.
In the previous section, we have proven that this scheme corresponds to 
a quasi-stationary case associated with high nucleation barriers.
The theory we propose also allows us to evaluate the evolution and the time-dependent
rate of change of the global crystallinity $Q$ of the system, at any intermediate
state between $Q_{1}$ and $Q_{2}$.

Our scheme is formulated in terms of a Kramers-like equation accounting for the
 evolution of $\rho(Q,t)$. From the knowledge of the number 
density $\rho(Q,t)$ 
one could then evaluate the relevant quantities of 
the system.

 In the general case, one must work with the kinetic equation (\ref{eq:FPQ})
  which constitutes the central point of our
approach. This equation contains the free energy barrier 
$\Phi(Q)$
and the phenomenological coefficient in the internal space $L(Q)$.
The barrier
 $\Phi(Q)$ could be evaluated from simulations analogous to the ones performed in 
reference
\ref{bib:barrera}. In section III we have obtained an explicit expression 
for the coefficient $L(Q)$. However, it is necessary to transform eq.
 (\ref{eq:gkq}) into an expression more proper for simulations.

With the help of eq. 
(\ref{eq:contideltaN}), one can relate the rate
 of change of the order parameter to
the random contribution to the current in the following way

\begin{equation}\label{eq:Qpuntodete} \langle \dot{Q} (t)\rangle \,=\,
\int_{Q_{1}}^{Q_{2}} Q \frac{\partial \delta \rho(Q,t)}{\partial t} dQ\,=\,
\int_{Q_{1}}^{Q_{2}}J^{r}(Q,t) dQ \end{equation}

\noindent By substitution of the latter expression and eq.  
(\ref{eq:contideltaN}) 
into the Green-Kubo formula (\ref{eq:gkq}) for $L(\gamma)$ (where $\gamma$ 
represents an arbitrary value of the degree of crystallization) we finally
 obtain

\begin{equation}\label{eq:LQgene} k_{B}L(\gamma)\,=\, -\int_{0}^{\infty}dt
 \langle { \langle \dot{Q}(0) \rangle \delta\dot{N}(\gamma,t) } \rangle_{eq}
 \end{equation}

\noindent   In the equation above there
 appears the
expected value $\langle \dot{Q} (0)\rangle$, which could be identified with the
initial speed of change $ \dot{Q} (0)$ of the system which is being simulated.
Additionally, the term $\delta\dot{N}(\gamma,t)$ is given by

\begin{equation}\label{eq:deNQ} \delta \dot{N}(\gamma,t)\,=\,- \langle
\dot{n}_{\gamma} (t) \rangle \end{equation}

\noindent where ${n}_{\gamma} (t)\,=\,\theta [ Q(t)\,-\, \gamma ]$.

With these considerations in mind, the definitive expression in a form 
suitable for simulation yields

\begin{equation}\label{eq:LQfinal} L(\gamma)\,=\, 
\frac{1}{k_{B}}\int_{0}^{\infty}dt
 \langle {  \dot{Q}(0) \dot{n}_{\gamma} (t)  } \rangle_{eq}
 \end{equation}

Note that the latter equation for the phenomenological coefficient 
is formally similar to expression
(\ref{eq:kabchandler}),
analyzed in  sections II and IV. However, two significant differences exist. 
The first difference comes from the fact that in eq. (\ref{eq:kabchandler}), the only 
trajectories giving a non-vanishing 
contribution are the ones corresponding to configurations initially
at the top of the barrier. The second consists of the replacement of the 
characteristic function
\mbox{ $n_{B}(t)\,=\,\theta [
 Q(t)\,-\,Q_{0} ]$ } of eq. (\ref{eq:kabchandler}) for $n_{\gamma}(t)$. 
Therefore, by performing simulations analogous to the ones carried out 
by Frenkel and co-workers, 
   we could obtain  the phenomenological coefficient $L(\gamma)$ for any value
$\gamma$ of the degree of crystallization. 
     Knowing $L(\gamma)$, since the barrier
 $\Phi(Q)$ could be evaluated from simulations, we could completely determine 
the 
Kramers-like eq. (\ref{eq:FPQ}).  By solving this 
equation we could determine
 the probability density $\rho(Q,t)$ describing the dynamics of the system.

\section{Conclusions}

In this paper we have introduced a systematic scheme to treat
homogeneous nucleation, applicable to any type of global order parameters
describing the process. This scheme is based on the internal degrees 
of freedom formalism inside the framework of non-equilibrium thermodynamics.
In fact, we have provided an extension of this formalism to allow 
descriptions in terms of order parameters,
relying on the construction of a non-equilibrium ensemble.

  Our treatment links theory with simulations,
in the sense that the equation we propose is a Kramers-like equation as the ones
commonly used at the theoretical level, but its coefficients are susceptible to be
determined from simulations. Moreover, it constitutes a unifying proposal where the expressions of the method of reactive
flux emerge in the quasi-stationary limit of a more general Kramers-type theory.

The approach we have introduced provides a complete and fully consistent 
description
of homogeneous nucleation for an order parameter such as the one used by Frenkel 
and co-workers in their simulations.  Our scheme not only includes the reactive flux 
theory,
but also it allows a deeper study of the process, offering a framework for a
theoretical interpretation of simulation results. We have outlined that 
 by performing simulations analogous to the ones carried out in the quasi-stationary
case, one could in principle obtain more complete information about
the process.  For example, we could study time-dependent nucleation and follow
the evolution and the rate of change of intermediates degrees of crystallinity
$Q$.

Additionally, our
scheme casts light on the drawbacks exposed by Frenkel $et\;al.$  Our picture 
overcomes the first objection, in the sense that our
equation constitutes the proof of the validity of a Kramers-like diffusive
equation for global order parameters.  The equation we propose also takes into
account the possible dependences on $Q$ and even on $t$ of the diffusion
coefficient.  Moreover, it gives explicit expressions to evaluate it, which in
 fact are very similar to the ones used
 by Frenkel and co-workers in their simulations.
 
Finally, with respect to their last concern we have to remark that the fact of
working with a global description, where one characterizes the
macroscopic state of the system by means of a single coordinate, implies that
hydrodynamic effects, that are of local character, cannot be taken into
account.  However, by employing a local description of the process (like the one
used in ref. \ref{bib:yo}), the formalism we have introduced enables us to study
hydrodynamic effects in nucleation processes.  In order to analyze these
effects, we must increase the number of 
variables to achieve a more detailed description of the system, and consider additional contributions in
the Gibbs eq. (\ref{eq:dentropy}).  The way to proceed has been outlined in 
references \ref{bib:agustin} and 
\ref{bib:rubi} and will be applied to nucleation processes in a future 
contribution.

\acknowledgments

We would like to thank Prof. H. Reiss and T. Alarc\'on for fruitful discussions. 
This work has been supported by DGICYT of the Spanish Government under
grant PB95-0881.  \mbox{D.  Reguera} wishes to thank Generalitat de
Catalunya for financial support.

\begin{center} BIBLIOGRAPHY \end{center}

\begin{enumerate}

\item \label{bib:degroot}S.  R. de Groot and P. Mazur, $Non\!-\!Equilibrium\; 
Thermodynamics $, (Dover, New York, 1984).

\item \label{bib:frenkel}P. R. ten Wolde, M. J. Ruiz-Montero and D. Frenkel.
 J. Chem. Phys. {\bf 104}, 9932 (1996).

\item \label{bib:frenkel2}P. R. ten Wolde, M. J. Ruiz-Montero and D. Frenkel.
 Faraday Discuss. {\bf 104}, 93 (1996).

\item \label{bib:frenkel3}P. R. ten Wolde, M. J. Ruiz-Montero and D. Frenkel.
 Phys. Rev. Lett. {\bf 75}, 2714 (1995).

\item \label{bib:reviews} Some reviews on nucleation:  a) K.  F.  Kelton.  Solid 
State
Phys.  {\bf 45}, 75 (1991).  b) D.  T.  Wu.  Solid State Phys.  {\bf 50}, 37
(1996).  c) A.  C.  Zettlemoyer.  $ Nucleation $.  (Marcel Dekker, New York,
1969).  d) A.  Laaksonen, V.  Talanquer and D.  W.  Oxtoby.  Annu.  Rev.  Phys.
Chem.  {\bf 46}, 489 (1995).  e) P.  G.  Debenedetti.  $Metastable \; Liquids\!:
concepts\; and\; principles$.  Chap.  3.  (Princeton University Press, 1996).

\item \label{bib:kramers}H. A. Kramers.
Physica  {\bf VII}, 284 (1940).

\item \label{bib:hanggi} P.  H\"{a}nggi, P. Talkner and M. Borkovec. 
Rev. Mod. Phys. {\bf 62}, 251 (1990).

\item \label{bib:simula}For example:
a) C. S. Hsu and A. Rahman.  J. Chem. Phys. {\bf 71}, 4974 (1979).
b) R. D. Mountain and A. C. Brown. J. Chem. Phys. {\bf 80}, 2730 (1984).
c) S. Nos\'e and F. Yonezawa. J. Chem. Phys. {\bf 84}, 1803 (1986).
d) L. A. B\'aez and P. Clancy. J. Chem. Phys. {\bf 102}, 8183 (1995).

\item \label{bib:barrera}J. S. van Duijneveldt and D. Frenkel.
J. Chem. Phys. {\bf 96}, 4655 (1992).

\item \label{bib:chandler}D.  Chandler.
J. Chem. Phys. {\bf 68}, 2959 (1978).

\item \label{bib:chandler2}D.  Chandler, $Introduction \;to\; Modern \;Stastistical\;
 Mechanics$, (Oxford University Press, 1987).

\item \label{bib:bennett}C.  H. Bennett, in $Algorithms \;for\; Chemical \;
Computation$, edited by R. E. Christofferson (American Chemical Society, 
Washington. D. C., 1977).

\item \label{bib:param} P. J. Steinhardt, D. R. Nelson and M. Ronchetti.
 Phys. Rev. B {\bf 28}, 784 (1983).

\item \label{bib:frenkel4} M. J. Ruiz-Montero, D. Frenkel. and J. J. Brey.
 Molecular Physics {\bf 90}, 925 (1997).

\item \label{bib:yo}D. Reguera, J. M.  Rub\'{\i} and A.  P\'erez-Madrid.
Physica A, in press.

\item \label{bib:prigo}I. Prigogine and P. Mazur.
Physica  {\bf XIX}, 241 (1953).

\item \label{bib:agustin}A.  P\'erez-Madrid, J. M.  Rub\'{\i}, and P. Mazur. 
Physica  A {\bf 212}, 231 (1994).

\item \label{bib:rubi} J. M.  Rub\'{\i} and P. Mazur. Physica A, 
{\bf 250}, 253 (1998).

\item \label{bib:absor}I.  Pagonabarraga and J. M.  Rub\'{\i}. 
Physica  A {\bf 188}, 553 (1992).

\item \label{bib:pago}I.  Pagonabarraga, A.  P\'erez-Madrid, J. M.  Rub\'{\i}. 
Physica  A {\bf 237}, 205 (1997).

\item \label{bib:gabriel}G.  Gomila, A.  P\'erez-Madrid, J. M.  Rub\'{\i}.
Physica  A {\bf 233}, 208 (1996).

\item \label{bib:landau}L. D.  Landau and E. M. Lifshitz, $Course\; of\; 
Theoretical\; 
Physics\; Vol \;5 \;$ (Statistical Physics Part 1) \S 118 and $Vol \;9 $ (Statistical Physics Part 2) (Pergamon Press, New York, 1980).

\item \label{bib:zwanzing}R.  Zwanzig. 
Annu. Rev. Phys. Chem. {\bf 16}, 67 (1965).

\end{enumerate}

\end{document}